\documentclass{article}
\usepackage[psamsfonts]{amssymb}
\usepackage{amsmath}
\usepackage{cite}

\usepackage{cite}
\def\rd{\mathrm{d}}
\def\ri{\mathrm{i}}
\author{Leila Ramezan\footnote{l.ramezan@gmail.com}\ \ and
Mohammad Khorrami\footnote{mamwad@mailaps.org}
\\ {\small Department of Physics, Alzahra University, Tehran
1993891167, Iran.}}
\title{Spin 0 and spin 1/2 particles in a spherically symmetric static gravity and a Coulomb field}
\date{}
\begin{document}
\maketitle

\noindent \textbf{keywords}: relativistic particle, gravitational field, energy levels


\begin{abstract}
\noindent A relativistic particle in an attractive Coulomb field as well
as a static and spherically symmetric gravitational field is
studied. The gravitational field is treated perturbatively
and the energy levels are obtained for both spin $0$ (Klein-Gordon)
and spin $1/2$ (Dirac) particles. The results are shown to coincide
with each other as well as the result of the nonrelativistic
(Schr\"{o}dinger) equation in the nonrelativistic limit.
\end{abstract}
\section{Introduction}
There has been a large collection of investigations on the systems in which
both quantum mechanics and gravity are important. Some of these studies are
on quantum fields in curved bakgrounds, (\cite{BD} for example). Some others
are on the behavior of first-quantized systems in curved backgrounds. Among
these one can mention the simple study of a nonrelativistic particle in the
presence of a constant gravity (\cite{LL}, for example), the effect of
gravity on neutrino oscillations (\cite{PRW,FGKS,AS}), and also the study of
relativistic quantum equations in a gravitational background. Chandrasekhar
has studied the Dirac equation in a Kerr background \cite{Cs}, and others
have extended this study in \cite{Cb,MC} (for example). People have also
studied the effect of spin. In \cite{VR} the possible difference between the
Dirac hamiltonians in a Schwarzschild background and a uniformly accelerated
frame and the role of spin in this difference is addressed. In \cite{KAS,AV},
the solutions of the klein-Gordon and Dirac equation in specific gravitational
backgrounds have been studied and their differences have been addressed.

Here we are going to investigate the energy levels corresponding to
bound states of charged particles in the field of a static charge in a curved
static rotationally symmetric background. The system is studied perturbatively.
The scheme of the paper is the following. In section 2 the form of the Klein-Gordon
and Dirac equation in a static rotationally symmetric background is reviewed. In
section 3, the energy levels corresponding to the Klein-Gordon and Dirac equation
are obtained up to first order in the deviation of the metric from the metric of
a flat space-time. In section 4 the nonrelativistic limit is investigated.
Section 5 is a discussion of the length scales involved, and section 6 is devoted to the
concluding remarks.

Throughout this text, the signature $(-+++)$ is used for the metric.

\section{The Klein-Gordon and Dirac equation in a static rotationally symmetric background}
To fix notation, let us quickly review the form of Klein-Gordon and Dirac equations.
The line element in suitable coordinates is
\begin{align}\label{cd.1}
\rd s^2&=-c^2\,A^2(r)\,\rd t^2+B^2(r)\,\rd r^2+r^2\,(\rd\theta^2+\sin^2\theta\,\rd\phi^2),\nonumber\\
&=:g_{\mu\,\nu}\,\rd x^\mu\,\rd x^\nu.
\end{align}
The Klein-Gordon equation in the presence of a vector potential is
\begin{align}\label{cd.2}
\Bigg[&\frac{1}{\sqrt{|\det g|}}\,\left(-\ri\,\hbar\,c\,\frac{\partial}{\partial x^\mu}-U_\mu\right)
\,\sqrt{|\det g|}\,g^{\mu\,\nu}\nonumber\\
&\times\left(-\ri\,\hbar\,c\,\frac{\partial}{\partial x^\nu}-U_\nu\right)
-m^2\,c^4\Bigg]\,\psi_\mathrm{KG}=0,
\end{align}
where $U$ is the vector potential times the particle's charge, and $m$ is the particle's mass. For
the line element (\ref{cd.1}), and if the only nonzero component of the vector potential is
the zeroth (time) component, one arrives at
\begin{equation}\label{cd.3}
\left[-\left(\ri\,\hbar\,\frac{\partial}{\partial t}-V\right)^2+
c^2\,A^2\,(-\hbar^2\,\nabla^2+m^2\,c^2)\right]\,\psi_\mathrm{KG}=0,
\end{equation}
where
\begin{equation}\label{cd.4}
\nabla^2:=\frac{1}{A\,B\,r^2}\,\frac{\partial}{\partial r}\,\frac{A\,r^2}{B}\,\frac{\partial}{\partial r} +\frac{1}{r^2}\,\nabla^2_\Omega,
\end{equation}
and
\begin{equation}\label{cd.5}
\nabla^2_\Omega:=\frac{1}{\sin\theta}\,\frac{\partial}{\partial\theta} \,\sin\theta\,\frac{\partial}{\partial\theta}+\frac{1}{\sin^2\theta}\,\frac{\partial^2}{\partial\phi^2}.
\end{equation}

The Dirac equation is
\begin{equation}\label{cd.6}
\{\gamma^a\,[\hbar\,c\,(\partial_a+\Gamma_a)-\ri\,U_a]-m\,c^2\}\,\psi_\mathrm{D}=0,
\end{equation}
where the indices $a$, $b$, ... denote tetrad indices, $\gamma^a$'s are the Dirac matrices satisfying
the Clifford algebra
\begin{equation}\label{cd.7}
\{\gamma^a,\gamma^b\}=2\,\eta^{a\,b},
\end{equation}
in which
\begin{equation}\label{cd.8}
\eta^{a\,b}:=\begin{cases}-c^{-2},&a=b=0\\ 1&a=b\ne 0\\ 0&a\ne b\end{cases},
\end{equation}
and $\Gamma_a$'s are spin connections.
These connections are found through
\begin{equation}\label{cd.9}
\Gamma_a:=-\frac{1}{8}\,[\gamma_b,\gamma_c]\,\Gamma^c{}_a{}^b,
\end{equation}
where $\Gamma^c{}_a{}^b$ is antisymmetric in $c$ and $b$ and satisfies
\begin{equation}\label{cd.10}
\rd e^c+\Gamma^c{}_{a\,b}\,e^a\wedge e^b=0,
\end{equation}
in which $e^a$'s form a tetrad basis:
\begin{equation}\label{cd.11}
e^a\cdot e^b=\eta^{a\,b}.
\end{equation}
choosing the tetrad
\begin{align}\label{cd.12}
e^0&:=A\,\rd t,\nonumber\\
e^1&:=B\,\rd r,\nonumber\\
e^2&:=r\,\rd\theta,\nonumber\\
e^3&:=r\,\sin\theta\,\rd\phi,
\end{align}
corresponding to the line element (\ref{cd.1}), and assuming that the only nonzero component of the
vector potential is the zeroth (time) component, one arrives at
\begin{align}\label{cd.13}
\bigg[&-\left(\ri\,\hbar\,\frac{\partial}{\partial t}-V\right)-\frac{\ri\,\hbar\,c\,A}{B}\,
\left(\frac{\partial}{\partial r}+\frac{1}{r}+\frac{1}{2\,A}\,\frac{\rd A}{\rd r}\right)\,\alpha^1
-\frac{\ri\,\hbar\,c\,A}{r}\,\alpha^1\,\beta\,\hat K\nonumber\\
&+m\,c^2\,A\,\beta\bigg]\,\psi_\mathrm{D}=0,
\end{align}
where
\begin{equation}\label{cd.14}
\hat K:=\beta\,\left[\alpha^1\,\alpha^2\,\left(\frac{\partial}{\partial\theta}+
\frac{1}{2}\,\cot\theta\right)+
\alpha^1\,\alpha^3\,\frac{1}{\sin\theta}\,\frac{\partial}{\partial\phi}\right],
\end{equation}
and
\begin{align}\label{cd.15}
\beta&:=\ri\,c\,\gamma^0,\nonumber\\
\alpha^j&:=c\,\gamma^0\,\gamma^j.
\end{align}
$\alpha^j$'s and $\beta$, of course satisfy
\begin{align}\label{cd.16}
\{\alpha^i,\alpha^j\}&=2\,\delta^{i\,j},\nonumber\\
\{\alpha^j,\beta\}&=0,\nonumber\\
\beta^2&=1.
\end{align}

Finally, for the potential energy corresponding to a point charge in the field of a point charge at
the origin, one has
\begin{equation}\label{cd.17}
\frac{\partial}{\partial x^j}(F^{j\,0}\,\sqrt{|\det g|})=0,\qquad r\ne 0,
\end{equation}
which (using spherical symmetry) results in
\begin{equation}\label{cd.18}
F^{r\,t}\,A\,B\,r^2=\mathrm{constant},
\end{equation}
or
\begin{equation}\label{cd.19}
F_{r\,t}\,(A\,B)^{-1}\,r^2=\mathrm{constant},
\end{equation}
where $F$ is the field strength tensor. Using these, one arrives at the following equation for $V$:
\begin{equation}\label{cd.20}
\frac{\rd V}{\rd r}=\frac{k\,A\,B}{r^2},
\end{equation}
where $k$ is a constant, actually minus the product of charges times the constant used in the
Coulomb force expression.

\section{Perturbative calculation of the energy levels}
Assume that the gravitational acceleration vanishes at $r=0$. This means that the first derivatives of
the metric vanish at the origin. Assuming that there is no singularity at $r=0$ (so that there is no
angle deficit), one arrives at
\begin{equation}\label{cd.21}
B(0)=1.
\end{equation}
By a suitable (constant) scaling of the time, one can also make
\begin{equation}\label{cd.22}
A(0)=1.
\end{equation}
So near $r=0$ one can expand $A$ and $B$ like
\begin{align}\label{cd.23}
A(r)&=1+\xi\,r^2,\nonumber\\
B(r)&=1+\upsilon\,r^2.
\end{align}
using these, up to first order in $\xi$ and $\upsilon$ one arrives at
\begin{equation}\label{cd.24}
V=V_0+k\,(\xi+\upsilon)\,r,
\end{equation}
where
\begin{equation}\label{cd.25}
V_0:=-\frac{k}{r}.
\end{equation}
\subsection{The Klein-Gordon equation}
Using (\ref{cd.4}), one has (up to first order)
\begin{equation}\label{cd.26}
\nabla^2=(1-2\,\upsilon\,r^2)\,\nabla_0^2+\left[2\,\upsilon\,\nabla_\Omega^2+
2\,(\xi-\upsilon)\,r\,\frac{\partial}{\partial r}\right],
\end{equation}
where
\begin{equation}\label{cd.27}
\nabla_0^2:=\frac{1}{r^2}\,\frac{\partial}{\partial r}\,r^2\,\frac{\partial}{\partial r} +\frac{1}{r^2}\,\nabla^2_\Omega.
\end{equation}
Letting $\psi_\mathrm{KG}$ be a common eigenfunction of $\ri\,\hbar\,\partial/\partial t$ and
$\nabla^2_\Omega$:
\begin{align}\label{cd.28}
\ri\,\hbar\,\frac{\partial}{\partial t}\psi_\mathrm{KG}&=E\,\psi_\mathrm{KG},\nonumber\\
\nabla^2_\Omega\,\psi_\mathrm{KG}&=\lambda\,\psi_\mathrm{KG},
\end{align}
where
\begin{equation}\label{cd.29}
\lambda=-\ell\,(\ell+1)
\end{equation}
and $\ell$ is a nonnegative integer, one arrives (up to first order)
\begin{align}\label{cd.30}
E^2\,\psi_\mathrm{KG}=\bigg\{&2\,E\,V-V^2+[1+2\,(\xi-\upsilon)\,r^2]\,c^2
\,(-\hbar^2\,\nabla_0^2+m^2\,c^2)\nonumber\\
&+2\,m^2\,c^4\,\upsilon\,r^2-\hbar^2\,c^2\,\left[2\,\upsilon\,\nabla_\Omega^2+
2\,(\xi-\upsilon)\,r\,\frac{\partial}{\partial r}\right]\bigg\}\,\psi_\mathrm{KG}.
\end{align}
Multiplying both sides by $\psi_{\mathrm{KG}\,0}$ (the unperturbed eigenfunction corresponding
to the unperturbed energy $E_0$), and normalizing $\psi_\mathrm{KG}$ like
\begin{equation}\label{cd.31}
\langle\psi_{\mathrm{KG}\,0},\psi_\mathrm{KG}\rangle=1,
\end{equation}
one arrives at
\begin{align}\label{cd.32}
E^2=&\,E_0^2+\langle(2\,E\,V-V^2-2\,E_0\,V_0+V_0^2)\rangle
+\langle2\,(\xi-\upsilon)\,r^2\,c^2\,
(-\hbar^2\,\nabla_0^2+m^2\,c^2)\rangle
\nonumber\\
&+\left\langle\bigg\{2\,m^2\,c^4\,\upsilon\,r^2-\hbar^2\,c^2\,
\left[+2\,\lambda\,\upsilon+2\,(\xi-\upsilon)\,r\,\frac{\partial}{\partial
r}\right]\bigg\}\right\rangle,
\end{align}
where expectation values are calculated with the unperturbed eigenfunction.
Finally, one has
\begin{align}\label{cd.33}
(E_0-\langle V_0\rangle)\,\Delta E
=&\,2\,k^2\,\xi-\hbar^2\,c^2\,\lambda\,\upsilon+E_0\,k\,(3\,\xi-\upsilon)\,\langle
r\rangle\nonumber\\
& +[E_0^2\,(\xi-\upsilon)+\,m^2\,c^4\,\upsilon]\,\langle r^2\rangle
-\hbar^2\,c^2\,(\xi-\upsilon)\,\left\langle r\,\frac{\partial}{\partial
r}\right\rangle,
\end{align}
where
\begin{equation}\label{cd.34}
\Delta E:=E-E_0.
\end{equation}
To obtain the expectation values, one notices that the unperturbed Klein-Gordon
equation is like a Schr\"{o}dinger equation with modified parameters:
\begin{equation}\label{cd.35}
\left(-\frac{\hbar^2}{2\,m}\,\nabla_0^2+\frac{E_0\,V_0}{m\,c^2}
-\frac{V_0^2}{2\,m\,c^2}\right)\,\psi_0=
\frac{E_0^2-m^2\,c^4}{2\,m\,c^2}\,\psi_0.
\end{equation}
So defining
\begin{align}\label{cd.36}
\epsilon_0&:=\frac{E_0^2-m^2\,c^4}{2\,m\,c^2},\nonumber\\
\tilde k&:=\frac{E_0\,k}{m\,c^2},\nonumber\\
s\,(s+1)&:=\ell\,(\ell+1)-\frac{k^2}{\hbar^2\,c^2},
\end{align}
one arrives at
\begin{equation}\label{cd.37}
\epsilon_0=-\frac{m\,c^2}{2}\,\frac{\tilde
k^2}{\hbar^2\,c^2}\,\frac{1}{(n'+s+1)^2},
\end{equation}
where $n'$ is a nonnegative integer. Then
\begin{equation}\label{cd.38}
E_0=m\,c^2\,\left[1+\frac{k^2}{\hbar^2\,c^2\,(n'+s+1)^2}\right]^{-1/2}.
\end{equation}
One can then change $(k,\ell)$ to $(\tilde k,s)$ in the expectation values corresponding to
the Schr\"{o}dinger equation, to obtain the expectation values corresponding to the Klein-Gordon equation.
One has (see for example \cite{Sa})
\begin{align}\label{cd.39}
\left\langle\frac{1}{r}\right\rangle&=
-\frac{2\,\epsilon_0}{\tilde k},\\ \label{cd.40}
\langle r\rangle&=a\,\frac{3\,(n'+s+1)^2-s\,(s+1)}{2},\\ \label{cd.41}
\langle r^2\rangle&=
a^2\,\frac{(n'+s+1)^2\,[5\,(n'+s+1)^2+1-3\,s\,(s+1)]}{2},
\end{align}
where
\begin{equation}\label{cd.42}
a:=-\frac{\tilde k}{2\,(n'+s+1)^2\,\epsilon_0}.
\end{equation}
Also, as $(1/r)(\partial/\partial r)\,r$ is antihermitian and $r$ is hermitian, one has
\begin{equation}\label{cd.43}
\left\langle r\,\frac{1}{r}\,\frac{\partial}{\partial r}\,r
+\frac{1}{r}\,\frac{\partial}{\partial r}\,r\,r\right\rangle=0,
\end{equation}
so that
\begin{equation}\label{cd.44}
\left\langle r\,\frac{\partial}{\partial
r}\right\rangle=-\frac{3}{2}.
\end{equation}
Putting all these together,
\begin{align}\label{cd.45}
\Delta E=\frac{E_0\,\hbar^2\,c^2}{m^2\,c^4}\,\bigg[&
\frac{3}{2}\,(\xi-\upsilon)+2\,\frac{k^2}{\hbar^2\,c^2}\,\xi+\ell\,(\ell+1)\,\upsilon\nonumber\\
&+(3\,\xi-\upsilon)\,\frac{3\,(n'+s+1)^2-s\,(s+1)}{2}\nonumber\\
&+\left(\frac{E_0^2}{m^2\,c^4-E_0^2}\,\xi+\upsilon\right)
\,\frac{5\,(n'+s+1)^2+1-3\,s\,(s+1)}{2}\bigg],
\end{align}
or
\begin{align}\label{cd.46}
\frac{\Delta E_\mathrm{KG}}{m\,c^2}=&\;f^2\,a_0^2\,
\left[1+\frac{f^2}{(n'+s+1)^2}\right]^{-1/2}\nonumber\\
&\times\bigg\{\frac{3}{2}\,(\xi-\upsilon)+2\,f^2\,\xi+\ell\,(\ell+1)\,\upsilon
\nonumber\\&\quad
+(3\,\xi-\upsilon)\,\frac{3\,(n'+s+1)^2-s\,(s+1)}{2}\nonumber\\
&\quad+\left[\frac{(n'+s+1)^2}{f^2}\,\xi+\upsilon\right]
\,\frac{5\,(n'+s+1)^2+1-3\,s\,(s+1)}{2}\bigg\},
\end{align}
where
\begin{align}\label{cd.47}
f&:=\frac{k}{\hbar\,c},\nonumber\\
a_0&:=\frac{\hbar^2}{m\,k}.
\end{align}
\subsection{The Dirac equation}
Using (\ref{cd.13}) and (\ref{cd.24}), one has (up to first order)
\begin{equation}\label{cd.48}
H_\mathrm{D}=[1+(\xi-\upsilon)]\,r^2]\,H_{\mathrm{D}\,0}+m\,c^2\,\upsilon\,r^2\,\beta+2\,k\,\xi\,r
-\ri\,\hbar\,c\,\hat K\,\upsilon\,r\,\alpha^1\,\beta-\ri\,\hbar\,c\,\xi\,r\,\alpha^1,
\end{equation}
where
\begin{equation}\label{cd.49}
H_{\mathrm{D}\,0}=V_0+m\,c^2\,\beta-\ri\,\hbar\,c\,
\left(\frac{\partial}{\partial r}+\frac{1}{r}\right)\,\alpha^1-
\frac{\ri\,\hbar\,c\,\hat K}{r}\,\alpha^1\,\beta.
\end{equation}
It is then seen that
\begin{equation}\label{cd.50}
\Delta E=(\xi-\upsilon)\,E_0\,\langle r^2\rangle+m\,c^2\,\upsilon\,\langle
r^2\,\beta\rangle+2\,k\,\xi\,\langle r\rangle-\ri\,\hbar\,c\,K\,\upsilon\,
\langle r\,\alpha^1\,\beta\rangle-\ri\,\hbar\,c\,\xi\,\langle r\,\alpha^1\rangle,
\end{equation}
where $E_0$ is the unperturbed energy, $K$ is the eigenvalue corresponding to $\hat K$, and the
expectation values are calculated using the unperturbed eigenvectors.

The expectation value of the commutator of anything with $H_{\mathrm{D}\,0}$
vanishes, in particular,
\begin{equation}\label{cd.51}
\langle[r^2,H_0]\rangle=0,
\end{equation}
which gives
\begin{equation}\label{cd.52}
\langle r\,\alpha^1\rangle=0;
\end{equation}
and
\begin{align}\label{cd.53}
\langle[r^w\,\beta,H_0]\rangle&=0,\nonumber\\
\langle[r^w\,\alpha^1,H_0]\rangle&=0,\nonumber\\
\langle[r^w\,\alpha^1\,\beta,H_0]\rangle&=0,\nonumber\\
\left\langle\left[\left(\frac{\partial}{\partial r}+\frac{1}{r}\right)\,r^w,H_0\right]\right\rangle&=0,
\end{align}
which give a set of recursive relations to obtain
\begin{align}\label{cd.54}
X_w&:=\langle r^w\rangle,\nonumber\\
Y_w&:=\langle r^w\,\beta\rangle,\nonumber\\
Z_w&:=\ri\,\hbar\,c\,\langle r^w\,\alpha^1\,\beta\rangle.
\end{align}
The relations are
\begin{align}\label{cd.55}
2\,E_0\,Y_w-2\,m\,c^2\,X_w+2\,k\,Y_{w-1}+w\,Z_{w-1}&=0,\nonumber\\
2\,m\,c^2\,Z_w-\hbar^2\,c^2\,w\,X_{w-1}+2\,\hbar^2\,c^2\,K\,Y_{w-1}&=0,\nonumber\\
2\,E_0\,Z_w+2\,k\,Z_{w-1}+2\,\hbar^2\,c^2\,K\,X_{w-1}-\hbar^2\,c^2\,w\,Y_{w-1}
&=0,\nonumber\\
-E_0\,(w+1)\,X_w+m\,c^2\,(w+1)\,Y_w-k\,w\,X_{w-1}-\,K\,w\,Z_{w-1}&=0.
\end{align}
One also has
\begin{align}\label{cd.56}
X_0&=1,\nonumber\\
Y_0&=\frac{1}{c^2}\,\frac{\partial E_0}{\partial m},\nonumber\\
&=\frac{E_0}{m\,c^2},
\end{align}
where
\begin{equation}\label{cd.57}
E_0=m\,c^2\,\left(1+\frac{k^2}{\hbar^2\,c^2\,\{n''+\sqrt{K^2-[k^2/(\hbar^2\,c^2)]}\}^2}\right)^{-1/2},
\end{equation}
and
\begin{equation}\label{cd.58}
|K|=j+\frac{1}{2},
\end{equation}
and $\hbar^2\,j\,(j+1)$ is the eigenvalue of the total angular momentum squared. These can be found
for example in \cite{ASa}.

Using (\ref{cd.55}) and (\ref{cd.56}), one arrives at
\begin{align}\label{cd.59}
Z_1=&\,\frac{\hbar^2\,c^2}{2\,m\,c^2}\,\left(1-\frac{2\,K\,E_0}{m\,c^2}\right),\nonumber\\
X_1=&\,\frac{k}{m^2\,c^4-E_0^2}\,\left(\frac{3\,E_0}{2}\right)
-\frac{\hbar^2\,c^2\,K}{2\,m\,c^2\,k}
-\frac{\hbar^2\,c^2\,K^2\,E_0}{2\,m^2\,c^4\,k},\nonumber\\
E_0\,X_2-m\,c^2\,Y_2=&\,-\frac{k^2\,E_0}{m^2\,c^4-E_0^2}+
\frac{\hbar^2\,c^2\,K^2\,E_0}{m^2\,c^4},\nonumber\\
X_2=&\,\frac{k^2}{(m^2\,c^4-E_0^2)^2}\,
\left(2\,E_0^2+\frac{m^2\,c^4}{2}\right)\nonumber\\
&\,+\frac{\hbar^2\,c^2}{m^2\,c^4-E_0^2}\,
\left(\frac{1-K^2}{2}-\frac{3\,K\,E_0}{2\,m\,c^2}
-\frac{K^2\,E_0^2}{m^2\,c^4}\right),
\end{align}
where
\begin{equation}\label{cd.60}
\Delta E=\xi\,E_0\,X_2-\upsilon\,(E_0\,X_2-m\,c^2\,Y_2)+2\,k\,\xi\,X_1-K\,\upsilon\,Z_1.
\end{equation}
So one has
\begin{align}\label{cd.61}
\frac{\Delta E_\mathrm{D}}{m\,c^2}=&\;\xi\,a_0^2\,\left[f^4\,
\frac{7\,\zeta^3-2\,\zeta}{2\,(\zeta^2-1)^2}-f^2\,
\frac{2\,K\,\zeta^2+(3\,K^2-1)\,\zeta+K}{2\,(\zeta^2-1)}\right]\nonumber\\
&\;+\upsilon\,a_0^2\,\left(f^4\,\frac{\zeta}{\zeta^2-1}-f^2\,
\frac{K}{2}\right),
\end{align}
where
\begin{equation}\label{cd.62}
\zeta:=\left[1+\frac{f^2}{(n''+\sqrt{K^2-f^2})^2}\right]^{1/2},
\end{equation}

\section{The nonrelativistic limit}
To obtain the nonrelativistic limit of the expressions (\ref{cd.46}) and (\ref{cd.61}),
one should take into account that $\xi$ and $\upsilon$ themselves are proportional to $c^{-2}$.
So expansion of the left hand sides of (\ref{cd.46}) and (\ref{cd.61}) up to order $f^2$ gives the
nonrelativistic value of $\Delta E$ and its first relativistic correction. Defining the principal quantum number through
\begin{equation}\label{cd.63}
n:=n'+\ell+1,
\end{equation}
in the Klein-Gordon equation and
\begin{equation}\label{cd.64}
n:=n''+|K|,
\end{equation}
in the Dirac equation, one arrives at
\begin{align}\label{cd.65}
\Delta E_\mathrm{KG}=m\,\bigg\{&(\xi\,c^2\,a_0^2)\,\frac{n^2\,[5\,n^2-3\,\ell\,(\ell+1)+1]}{2}\nonumber\\
&+(\xi\,c^2\,a_0^2)\,f^2\,\bigg[-\frac{10\,n^3+n}{2\,l+1}+\frac{19\,n^2+5}{4}
-\frac{3\,\ell\,(\ell+1)}{4}\nonumber\\
&\qquad+\frac{3\,n\,\ell\,(\ell+1)}{2\,\ell+1}\bigg]+(\upsilon\,c^2\,a_0^2)\,f^2\,(n^2-1)\bigg\}
+o(f^2),
\end{align}
and
\begin{align}\label{cd.66}
\Delta E_\mathrm{D}=m\,\bigg\{&(\xi\,c^2\,a_0^2)\,\frac{n^2\,[5\,n^2-3\,K\,(K+1)+1]}{2}\nonumber\\
&+(\xi\,c^2\,a_0^2)\,f^2\,\bigg[-\frac{10\,n^3+n}{2\,j+1}+\frac{19\,n^2}{4}+\frac{3\,n\,K}{2\,j+1}
+\frac{(2\,j+1)^2+4}{16}\nonumber\\
&\qquad-K\,(K+1)+\frac{3\,n\,(2\,j+1)}{4}\bigg]\nonumber\\
&+(\upsilon\,c^2\,a_0^2)\,f^2\,\bigg[n^2-\frac{K}{2}\bigg]\bigg\}+o(f^2),
\end{align}
where (\ref{cd.58}) has been used. In particular it is seen that in the limit $c\to\infty$,
the perturbed energy is the same for the Klein-Gordon and Dirac equation:
\begin{equation}\label{cd.67}
\Delta E_\mathrm{NR}=m\,(\xi\,c^2\,a_0^2)\,\frac{n^2\,[5\,n^2-3\,\ell\,(\ell+1)+1]}{2},
\end{equation}
where
\begin{equation}\label{cd.68}
K^2+\beta\,K=\frac{1}{\hbar^2}\,\mathbf{L}\cdot\mathbf{L}
\end{equation}
has been used (in which $\mathbf{L}$ is the orbital angular momentum), and the fact that
\begin{equation}\label{cd.69}
\lim_{c\to\infty}\langle\beta-1\rangle=0.
\end{equation}
Equation (\ref{cd.67}) is equivalent to
\begin{equation}\label{cd.70}
\Delta E_\mathrm{NR}=m\,(\xi\,c^2)\,\langle r^2\rangle_\mathrm{NR},
\end{equation}
which is nothing but the expectation value of the gravitational potential, as in the nonrelativistic limit
the only relevant term in the metric is the potential which is related to $A$ through
\begin{equation}\label{cd.71}
\left[1+\frac{\phi_\mathrm{gr}}{c^2}\right]^2=A^2.
\end{equation}

\section{Length scales}
It is seen from previous sections, eqs. (\ref{cd.65}) and (\ref{cd.66}), that the shift in the energy
levels up to leading order in $f$ is of order $m\,c^2\,\xi\,a_0^2$ and $m\,c^2\,\upsilon\,a_0^2\,f^2$.
Assuming that $f$ is small and $\xi$ and $\upsilon$ are of the same order $L^{-2}$, where $L^{-2}$ is
of the order of the curvature corresponding to the metric, it is seen that
\begin{equation}\label{cd.72}
\Delta E\sim m\,c^2\,\left(\frac{a_0}{L}\right)^2,
\end{equation}
or
\begin{equation}\label{cd.73}
\Delta E\sim (m\,c^2\,f^2)\,\left(\frac{a_0}{f\,L}\right)^2.
\end{equation}
In order that the (\ref{cd.23}) be valid, $\xi\,a_0^2$ and $\upsilon\,a_0^2$ should be
much less than unit, that is
\begin{equation}\label{cd.74}
a_0\ll L.
\end{equation}
In order that treating the gravitational field as a perturbation to the Coulomb field be valid,
the energy shift due to gravity should be much less than the unperturbed energy levels (minus the
rest energy of the particle of course). This reads
\begin{equation}\label{cd.75}
a_0\ll f\,L.
\end{equation}
Comparing (\ref{cd.74}) and (\ref{cd.75}), it is seen that for small Coulomb couplings (small
values of $f$), it is (\ref{cd.75}) which is more restrictive. So the overall criterion is
(\ref{cd.75}). If this is satisfied, then the transformation of the metric to something like a
Schwarzschild solution occurs at a length scale larger than the size of the system ($a_0$) divided
by $f$, which is much larger than the size of the system. So the form (\ref{cd.23}) is valid for the
system.

For a hydrogen atom, this means that the length scale corresponding to the
gravitational field should be much larger than a hundred Bohr radiuses. That is, $L$ should be
much larger than $10^{-8}$~m.

\section{Concluding remarks}
A static spherically symmetric gravitational field which vanishes at the origin is
determined by two constants near the origin. A particle in a Coulomb field as well as
such a gravitational field was studied. If the gravitational field is not very strong
so that its corresponding inverse curvature is much larger than the size of
the bounded system squared, it is plausible to treat the gravitational field
perturbatively. This was done and the perturbed energy levels were obtained for
spin $0$ and spin $1/2$ particles. The results were further expanded in terms of the
inverse of the speed of light, to deduce the nonrelativistic parts as well as the
first relativistic corrections. It was seen, as expected, that in the nonrelativistic
limit the shift of energy levels due to gravitation for the spin $0$ and spin $1/2$
particles coincide, and both coincide with the result of the Schr\"{o}dinger equation.
In fact the effect of the spin on the energy shift vanishes in this limit. In this limit
also only one of the parameters of the background metric enters the energy shift, the
parameter which determines the time-time component of the metric. This too, is expected,
as in the nonrelativistic limit only this parameter (which determines the nonrelativistic
gravitational potential) enters the equation of motion of the particle. Of course the
relativistic corrections contain the other parameter as well as spin-dependent terms.

\end{document}